# TruVR: Trustworthy Cybersickness Detection using Explainable Machine Learning

Ripan Kumar Kundu *
University of Missouri-Columbia

Rifatul Islam †
Northeastern University

Prasad Calyam ‡
University of Missouri-Columbia

Khaza Anuarul Hoque §
University of Missouri-Columbia

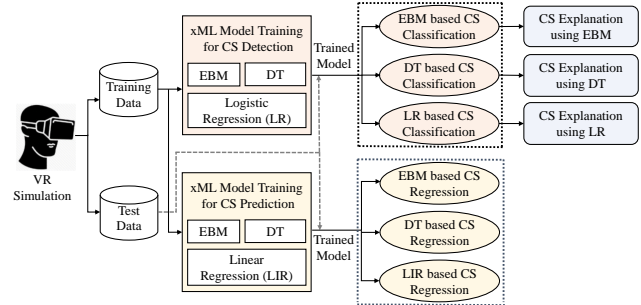

Figure 1: Overview of xML-based cybersickness (CS) classification, regression, and explanation framework for virtual reality applications.

## ABSTRACT

Cybersickness can be characterized by nausea, vertigo, headache, eye strain, and other discomforts when using virtual reality (VR) systems. The previously reported machine learning (ML) and deep learning (DL) algorithms for detecting (classification) and predicting (regression) VR cybersickness use black-box models; thus, they lack explainability. Moreover, VR sensors generate a massive amount of data, resulting in complex and large models. Therefore, having *inherent explainability* in cybersickness detection models can significantly improve the model's trustworthiness and provide insight into why and how the ML/DL model arrived at a specific decision. To address this issue, we present three explainable machine learning (xML) models to detect and predict cybersickness: 1) explainable boosting machine (EBM), 2) decision tree (DT), and 3) logistic regression (LR). We evaluate xML-based models with publicly available physiological and gameplay datasets for cybersickness. The results show that the EBM can detect cybersickness with an accuracy of 99.75% and 94.10% for the physiological and gameplay datasets, respectively. On the other hand, while predicting the cybersickness, EBM resulted in a Root Mean Square Error (RMSE) of 0.071 for the physiological dataset and 0.27 for the gameplay dataset. Furthermore, the EBM-based global explanation reveals exposure length, rotation, and acceleration as key features causing cybersickness in the gameplay dataset. In contrast, galvanic skin responses and heart rate are most significant in the physiological dataset. Our results also suggest that EBM-based local explanation can identify cybersickness-causing factors for individual samples. We believe the proposed xML-based cybersickness detection method can help future researchers understand, analyze, and design simpler cybersickness detection and reduction models.

**Index Terms:** Human-centered computing—Human computer interaction (HCI)—Interaction paradigms—Virtual reality; Human-centered computing—Human computer interaction (HCI)—HCI design and evaluation methods;

## 1 INTRODUCTION

Virtual reality (VR) has gained immense popularity in recent years and has been adapted in a wide variety of applications including medical training [29], education [48], national defense [3], public safety/disaster management [42] and many more [9, 17, 53, 59, 74]. However, VR users often experience **cybersickness** which is a set of unpleasant symptoms such as eyestrain, headache, nausea, disorientation, and such [10, 20, 48, 53, 59, 65, 66] which pose a serious threat to the immersive experience of the user. Hence, it is vital to detect and reduce cybersickness during VR immersion [8]. For example, it is feasible to apply cybersickness reduction techniques based on the severity of cybersickness [26] or to allow the player to evaluate their cybersickness score before purchasing a new VR game based on their projected cybersickness [1].

Motivated by this, recent works in cybersickness detection include the use of Deep Learning (DL) and Machine Learning (ML) from physiological signals (e.g., Heart Rate (HR), Galvanic Skin Responses (GSR), Breathe Rate (BR), Electroencephalogram (EEG)) [17, 50, 57, 68, 77] and using stereoscopic video [27, 41, 52]. For example, Islam et al. [27], Lee et al. [41], and Jeong et al. [30] used a complex multimodal deep fusion network with a 3D-CNN neural network and CNN-LSTM neural networks and found that features such as latency, optical-flow, disparity, saliency features, physiological signals (e.g., HR, GSR, etc.) are most prevalent features for causing cybersickness [13, 27, 28, 52]. However, when utilizing complex ML/DL models and fusing multimodal input from internal and external sensors, the black-box ML/DL models become increasingly complex, and their characteristics are difficult to interpret; thus, they lack explainability. Therefore, having *inherent interpretability* (i.e., without using any external explanation tools/methods) in ML/DL models can significantly improve the model's trustworthiness and provide insight into why and how the ML/DL model arrived at a specific decision. Indeed, understanding why some samples were incorrectly labeled as cybersickness is the first step toward discovering which feature contributed to the classification result.

We believe that understanding the feature spaces and contributing elements for cybersickness detection using explainable machine learning(xML)-based techniques will enable future research to create more effective cybersickness detection models. Therefore, in this research, we present **TruVR** – a framework for developing a trustworthy cybersickness detection method based on xML [43]. The *trustworthiness* in our proposed method is obtained via using a set of mechanisms, such as global and local explanation, and explainable layers to make the model transparent, understandable, and therefore, trusted by users [34, 69]. **Specifically, the contribution of this paper can be summarized as follows.**

- We exploit the inherent interpretability of Explainable Boosting Machine (EBM) [51], Decision Tree (DT) [21], Logistic Regression (LR) [39], and Linear Regression (LIR) [4] model

*e-mail: rkcgc@umsystem.edu
†e-mail: r.islam@northeastern.edu
‡e-mail: calyamp@umsystem.edu
§e-mail: hoquek@umsystem.edu

to classify and regress the cybersickness using two open-source datasets [24, 56].

- We also provide a global explanation (i.e., identifying features crucial for the overall prediction) and local explanation (i.e., identifying features dominating for an individual sample prediction) details for analyzing and identifying dominating features causing cybersickness.

It is worth mentioning that EBM is a tree-based, cyclic gradient boosting Generalized Additive Model with automatic interaction detection, which is often as accurate as state-of-the-art black box models while remaining completely interpretable [51]. EBMs are also knowns for being highly compact and fast at prediction time. All the utilized models in this work are inherently interpretable; thus, they do not require any external explanation tools, such as SHapley Additive exPlanations (SHAP) [5] for explaining the cybersickness detection outcomes. Our results show that the EBM model is highly accurate when compared to the DT, LR, and LIR models and classifies cybersickness (i.e., binary classification) with an accuracy of 99.75% for the physiological [24] dataset and 94.10% for the gameplay [56] dataset. In addition, the EBM model regresses ongoing cybersickness(i.e., Fast Motion Scale (FMS) score [35]) with a Root Mean Square Error (RMSE) 0.071 for the physiological [24] dataset and 0.27 for the gameplay [56] dataset.

## 2 RELATED WORKS

The most popular theories for cybersickness are the sensory conflict theory, poison theory, and postural instability theory [40]. Among them, the *sensory conflict theory* is the widely accepted one. According to the sensory conflict theory, cybersickness is caused by perceived pseudo-movement detected by visual sensory while the individual remains stationary [40]. Interestingly, cyberattack was also reported was one of the indirect reasons behind cybersickness in [22, 71, 72]. Additionally, there are certain variables that may influence the degree of cybersickness based on the participants' age, gender, and prior VR experience [14, 19, 23, 60]. In order to measure cybersickness researchers have proposed several subjective measurements such as the Simulator Sickness Questionnaire (SSQ) [10–12, 15, 64, 74], the FMS [35] and the Motion Sickness Susceptibility Questionnaire (MSSQ) [33]. In addition with subjective measures, objective measurements (i.e., physiological signals) for cybersickness have also been proposed by researchers [25, 28, 55]. According to previous research, objective measurement such as HR, GSR, and EEG vary significantly when cybersickness occurs. [13, 28, 44, 44, 61]. They found that HR and EEG delta waves positively correlate with cybersickness, whereas EEG beta waves have a negative correlation [44]. Another study reported that GSR has a much more positive correlation with cybersickness than other objective measurements and could be used to detect cybersickness [28, 67].

Recently, numerous ML and DL-based techniques have been proposed to automatically detect cybersickness from various objective and subjective data [2, 16, 25, 28, 31, 32, 36–38, 41, 73, 75]. For instance, Jin et al. [32] used three DL/ML algorithms: Convolutional neural network (CNN), Long short-term memory (LSTM), and support vector regression classifiers (SVM) to estimate the level of discomfort, where LSTM achieved the best results. Agundez et al. [18], utilizing a mix of physiological and game parameters along with users' respiratory and skin conductivity to analyze them using SVM and K-nearest neighbors (KNN) classifiers to classify the cybersickness severity. In [52], the authors used depth and optical flow features from the VR video data to predict cybersickness. In contrast, Lee et al. in [41] used a 3D-CNN and a multi-modal deep fusion approach with optical-flow, disparity, and saliency features and reported an improved accuracy for cybersickness detection when compared to the work in [52]. In [37], the authors used CNN and LSTM models to estimate the cognitive state using brain signals and how they relate to cybersickness levels. In contrast, the authors in [38] applied LSTM and Kalman filtering techniques. Authors in [28] classify the cybersickness severity from users' physiological signals (e.g. HR, GSR, etc.). In addition, a symbolic ML-based approach is presented in [55] to identify the levels of cybersickness.

Methods based on DL are excellent at detecting cybersickness. Nonetheless, the forecast conclusions are incomprehensible. As interpretability is crucial to comprehending cybersickness causes/symptoms, the fundamental complexity of DL models may restrict their utility. In this context, an xML-based approach can be advantageous. The application of xML has already been investigated in healthcare [45, 46, 63], finance [7], and law [62]. Sarica et al. used an EBM-based XML approach for predicting Alzheimer's disease from MRI Hippocampal Subfields [63].

Indeed, it is crucial to identify the key features inducing cybersickness in VR to develop effective mitigation methods [70] which can be achieved by using xML techniques. *However, to the best of our knowledge xML techniques for detecting and predicting cybersickness has not been explored yet, which motivates our work in the paper.*

## 3 METHODOLOGY

An overview of the proposed xML-based framework for cybersickness classification, regression, and explanation framework for VR applications (Figure 1). First, VR simulation data is divided into 70% for training and the remaining 30% for testing. Then, training data is used for training xML-based models (e.g., EBM, DT, and LR) for cybersickness classification and regression. Next, the trained xML-based classification and regression models are used for classifying and regressing the cybersickness from the test dataset. The cybersickness detection phase aims at classifying the test data in *cybersickness* and *no cybersickness* classes using the trained xML models. After the cybersickness classification phase, the cybersickness explanation is applied using global and local explanation methods. In a global explanation, the overall feature ranking is performed based on the overall outcome. On the contrary, the local explanation performs individual feature ranking based on the individual prediction. Finally, in the cybersickness prediction (regression) phase, the xML models regress the next value of the ongoing cybersickness FMS score range of 0 to 10. In the following sections, we describe the details of cybersickness classification, regression, and explanation.

### 3.1 Cybersickness Classification using EBM

EBM is built upon Generalized additive models (GAMs) [51]. Let us assume a cybersickness dataset $D$ contains a total $N$ number of samples, and $\beta$ denotes the learning intercept, $x_n$ denotes the feature from dataset $D$, $y_n$ represents the target label (i.e., cybersickness and no cybersickness), and $g_n$ denotes the non-linear functions namely, shape or feature function that describe the relationship between the output and input variable. Then, the link function that adapts the GAM for classification can be represented as: $f(E[y]) = \beta_0 + \sum g_n(x_n)$. To ignore the feature order, the boosting tree is applied in a round-robin fashion for each feature with a very low learning rate of 0.01. To learn the shape function $g_n$ for each feature cyclic gradient boosting [51] is considered with a number of iterations. To minimize the co-linearity, EBM iterates through each feature to learn the best feature function $g_n$ and provides the list of features that contribute to the prediction. The contribution of every feature to the final prediction can be visualized and understood by plotting the function $g_n$.

### 3.2 Cybersickness Classification using Decision Tree

DT constructs a tree-structured model to predict cybersickness outcomes based on the input features of the cybersickness dataset $D$.

Given samples $M$ from $D$, the expected information $Info(M)$ needed to correctly identify the cybersickness is given as:

$$Info(M) = -\sum_{i=1}^{m} p_i log_2(p_i) \quad (1)$$

In Equation 1, $p_i$ represents the probability of the cybersickness prediction that belongs to an actual cybersickness class $C_i$. The difference between the original information required and new information needed to predict cybersickness is defined as information gain $Gain(A) = Info(M) - Info_A(M)$, where $Info_A(M)$ denotes the expected information needed after partitioning using feature $A$ from the cybersickness dataset $D$. The gain ratio is then calculated [21]. In this paper, we first create the trees with a maximum depth of 3. Then, the attributes for the root node are identified. Next, the Gini cost function [21] is used to determine how pure a node is to select the next node. After prediction of the cybersickness, we provide the feature importance in terms of explanation.

### 3.3 Cybersickness Classification using Logistic Regression

This paper uses LR to predict cybersickness and then applies the cybersickness explanation. LR model has a fast training time, and it also provides good interpretability [6]. For the LR, we used the cross-entropy [39] loss function to measure the performance of a prediction model whose output has a probability value for cybersickness. Then, we explain the predicted cybersickness.

### 3.4 Cybersickness Regression with xML Models

In the regression analysis, we use three xML-based models, namely EBM, DT, and LIR, to estimate the cybersickness FMS score from 0 to 10. The cybersickness regression task can be defined as follows: Given the cybersickness, FMS score at previous time steps $t-1$ then, we have to predict the FMS score at the next time steps $t$. The predicted cybersickness FMS score at time $t$ based on the last time $t-1$ of physiological signals is denoted by $CS_t$. For instance, if we predict the cybersickness FMS score at time $t = 5$ seconds then, $CS_t$ can be written as: $CS_t \Rightarrow [M_{t-4}, M_{t-3}, M_{t-2}, \ldots, M_t]$, where $M$ denotes the user's physiological state.

### 3.5 Cybersickness Explanation with xML

The explainable ML methods produce visual representations, either as a DT (CART), scoring table (RiskSLIM) or as a set of visualizations either in global or local explanation [51]. We present these tables and visualizations for EBM, LR, and DT to give a clear understanding of each model's interpretability regarding cybersickness classification. We provide two explanations of the model decisions regarding global and local explanations. The overall feature importance ranking (global explanation) of cybersickness classification is visualized in bar graphs or CART. For the local explanation, each sample is randomly chosen from the test dataset, which contains all the features. For the local and global explanation, the mean absolute score (MAS) is used to calculate the feature importance during cybersickness classification. MAS is actually calculated as logits or log odds [79]. To convert these logits into a probability, we sum them up and pass them through the logistic link function. This logistic link function $L$ is calculated for the feature ranking as follows: $L(x) = \frac{1}{1+e^{-sum(logits)}}$, where $logits = x - x_0$ and $x_0$ represents the sigmoid's midpoint for the sample $x$.

## 4 DATASET & EXPERIMENTAL SETUP

This section explains the experimental setup and data used to validate and explain the xML-based classification and regression approaches for cybersickness. We used Python, and the Scikit-learn [54] for training and evaluating our xML models. For explaining the xML models, we used the InterpretML [51] library.

### 4.1 Datasets

To validate the effectiveness of the proposed xML models, we used the two datasets, such as gameplay [56] and physiological [24] datasets. The gameplay dataset [56] is comprised of 22 different features from the sources such as candidate profiles, questionnaires, user field of view, user position, speed of the game in playtime, etc. This dataset is generated using two VR games, i.e., racing and flight games, with a total of 88 participants. However, the data from 37 participants was stored due to their valid cybersickness. It is worth mentioning that the gameplay dataset contains a total of 9391 samples recorded with 5 minutes of VR gameplay simulation [55].

On the other hand, the physiological dataset [24] contains the physiological signals (such as HR, HR Variability (HRV), BR, and GSR) of 31 participants immersed in a VR roller coaster simulation. The HR, BR, GSR, and HRV data consist of four subcategories, including the percentage of change from resting baseline (PC), minimum inside 3s rolling window (MIN), the maximum value of 3s rolling window (MAX), and moving average of 3s rolling window (AVG). This dataset has a total of 14775 samples recorded with a maximum of 897 seconds of VR simulation. We applied the random oversampling [78] technique to the training samples of the physiological dataset due to class imbalance problems. As mentioned earlier, we used the 70% samples from both datasets for training the xML models and their remaining 30% samples for cybersickness classification and regression.

### 4.2 Ground-Truth Construction

In this paper, we proposed a binary classification model for cybersickness classification. Thus, in both datasets, the presence of cybersickness is labeled as 1, whereas the absence of cybersickness (no cybersickness) is labeled as 0. It is worth mentioning that the physiological dataset contains three different cybersickness severity classes: *low sickness*, *moderate sickness*, and *acute sickness*. We labeled 'moderate sickness' and 'acute sickness' as 1 and low sickness as 0 for this work. Similarly, for the gameplay dataset, the four original cybersickness classes are: *none*, *slight*, *moderate*, and *severe*, and we converted them into binary classes. We labeled 'none' as class 0 and 'slight', 'moderate', and 'severe' as class 1. For the regression analysis, the ground truth of the FMS score is labeled, ranging from 0 to 10 for both physiological and gameplay datasets.

### 4.3 Performance Metrics

The performance of the xML models for the cybersickness classification can be quantified using the standard quality metrics such as accuracy, precision, recall, F-1 score, the Area Under the Curve (AUC), and Receiver Operating Characteristic curve (ROC) [2]. Likewise, the performance of the regression models can be analyzed using the well-known loss functions such as Mean Square Error (MSE), RMSE, Mean Absolute Error (MAE), and $R^2$ score [27, 38, 52].

### 4.4 Hyper-Parameter

For training the EBM model, we used the learning rate of 0.001 for boosting the tree. To prevent the model from overfitting, we deployed an early-stopping strategy with a patience value of 30 while training the EBM model [51]. We used the default values for the rest of the parameters (e.g., maximum tree leaves, maximum bins, etc.). For training DT, we used trees with a maximum depth of 3 to prevent the model from overfitting [76]. Consequently, $L2$ penalty term is used for training the LR with binary cross-entropy loss function [39].

## 5 RESULTS

This section presents the results obtained from the cybersickness classification and regression using the EBM, LR, DT, and LIR-based methods.

Table 1: Cybersickness classification using xML models for the gameplay dataset

| xML models | Precision% | Recall% | F1-Score% | Acc.% |
|---|---|---|---|---|
| EBM | 87 | 90 | 88.94 | 94.10 |
| LR | 75 | 88 | 78.69 | 84.90 |
| DT | 72 | 85 | 74.14 | 80.51 |

Table 2: Cybersickness classification using xML models for the physiological dataset

| xML models | Precision% | Recall% | F1-Score% | Acc.% |
|---|---|---|---|---|
| EBM | 98.22 | 100 | 98.89 | 99.75 |
| LR | 72 | 75 | 73 | 77.92 |
| DT | 75 | 80 | 75 | 75.42 |

Table 3: Cybersickness regression using xML models for the physiological dataset

| xML Models | MSE | RMSE | $R^2$ | MAE |
|---|---|---|---|---|
| EBM | 0.005 | 0.071 | 0.975 | 0.0454 |
| LIR | 0.222 | 0.471 | -0.06 | 0.461 |
| DT | 0.1702 | 0.4126 | 0.19 | 0.2970 |

Table 4: Cybersickness regression results using xML models for the gameplay dataset

| xML Models | MSE | RMSE | $R^2$ | MAE |
|---|---|---|---|---|
| EBM | 0.073 | 0.27 | 0.45 | 0.071 |
| LIR | 0.1414 | 0.38 | 0.081 | 0.301 |
| DT | 0.127 | 0.372 | 0.41 | 0.197 |

### 5.1 Cybersickness Classification Using Decision Tree

Table 1 and Table 2 summarize the precision, recall, F-1 scores, and the accuracy of cybersickness classification using the DT model for both the gameplay and physiological datasets. The cybersickness classification using the DT model results in an accuracy of 80.51% for the gameplay dataset and 75.42% for the physiological dataset. The gameplay dataset's precision, recall, and F-1 score are 72%, 85% and 74.14%, and 75%, 80%, and 75% for the physiological dataset, respectively.

Although these results for both datasets seem close to each other considering their performance metrics, the AUC score in the case of the physiological dataset is 1.12 times less than that of the gameplay dataset. Hence, overall the gameplay dataset performs comparatively better than the physiological dataset in cybersickness classification. This is because the size of the physiological dataset is quite large. Therefore, a single tree may grow complex and cause the overfitting of the DT model. Due to overfitting, the variance in the output increases, leading to less efficient cybersickness classification.

### 5.2 Cybersickness Classification Using Logistic Regression

As shown in Table 1 and Table 2, the precision, recall, F1-score, and accuracy for cybersickness classification using LR are 75%, 88%, 78.69%, and 84.90% with the gameplay dataset and 72%, 75% and 73% and 77.92% with the physiological dataset, respectively. Figure 2 presents the AUC-ROC curves for both datasets (the dashed-dotted line represents the ROC values). The gameplay dataset possesses a higher AUC score of 0.887; however, the physiological dataset has a comparatively lower AUC score of 0.748. Hence, cybersickness classification using the LR model has better performance in the case of the gameplay dataset as compared to the physiological dataset. This is because the gameplay dataset contains features from mostly users' profile data, which is relatively linear; however, the physiological dataset is mostly non-linear. The LR lacks the capability of solving non-linear problems. Also, the physiological dataset's dimensionality is quite large compared to the gameplay dataset, which eventually leads to an over-fitting problem in LR.

### 5.3 Cybersickness Classification Using EBM

We observe that the cybersickness classification using EBM performs better than DT and LR models in the gameplay and the physiological datasets, as shown in Table 1 and Table 2. For instance, cybersickness classification using EBM for the gameplay dataset exhibits 94.10% accuracy, which is almost 14 and 9 times higher than that of DT and LR models. Likewise, cybersickness classification using EBM has 99.75% accuracy with the physiological dataset that is almost 24 and 22 times greater than that of DT and LR models, respectively. In addition, EBM achieves higher precision, recall, and F1-score for both datasets than other ML models. The precision, recall, and F1-score for the gameplay dataset are 87%, 90% and 88.94% and, 98.22%, 100% and 98.89% for the physiological dataset, respectively. Moreover, the AUC-ROC score for the cybersickness classification using EBM in Figure 2 is higher than that of DT and LR models in both datasets. The physiological dataset has an AUC score of 0.996, and the gameplay dataset has an AUC score of 0.913. Hence, EBM performs better than DT and LR models. This is because EBM assigns a tree for each feature, unlike DT and LR models, and then computes the probability. This provides depth insight into the feature, leading to high classification accuracy.

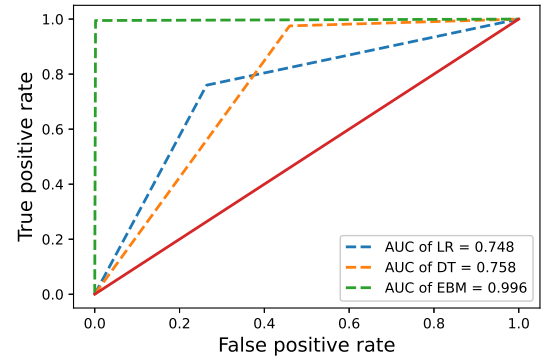

(a)

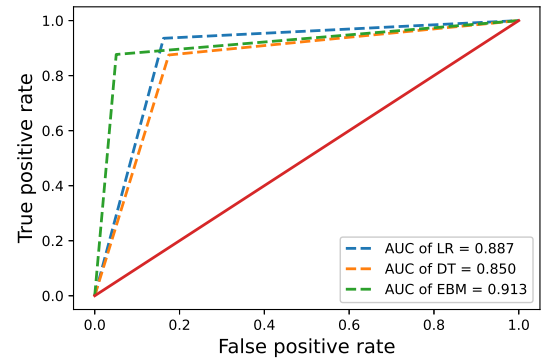

(b)

Figure 2: AUC-ROC curve for EBM, LR, and DT. (**a**) physiological dataset (**b**) gameplay dataset.

### 5.4 Cybersickness Regression Using Decision Tree

Table 3 and Table 4 shows the MSE, RMSE, MAE, and ($R^2$) values for the cybersickness regression using the DT model for both physiological and gameplay datasets. The MSE, RMSE, MAE and $R^2$ score values for the gameplay dataset are 0.127, 0.372, 0.197, and 0.41 respectively. However, the physiological dataset has comparatively higher MSE, RMSE, and MAE values and lower $R^2$ scores such as 0.1702, 0.4126, 0.2970, and 0.19 respectively.

Hence, cybersickness regression using the DT model has low performance with the physiological dataset compared to the gameplay dataset. The reason is that the DT does not perform well with continuous numerical variables. However, all measurements in the physiological dataset are continuous. Furthermore, unlike the physiological dataset, the gameplay dataset contains the user's profile data, which belongs to categorical variables.

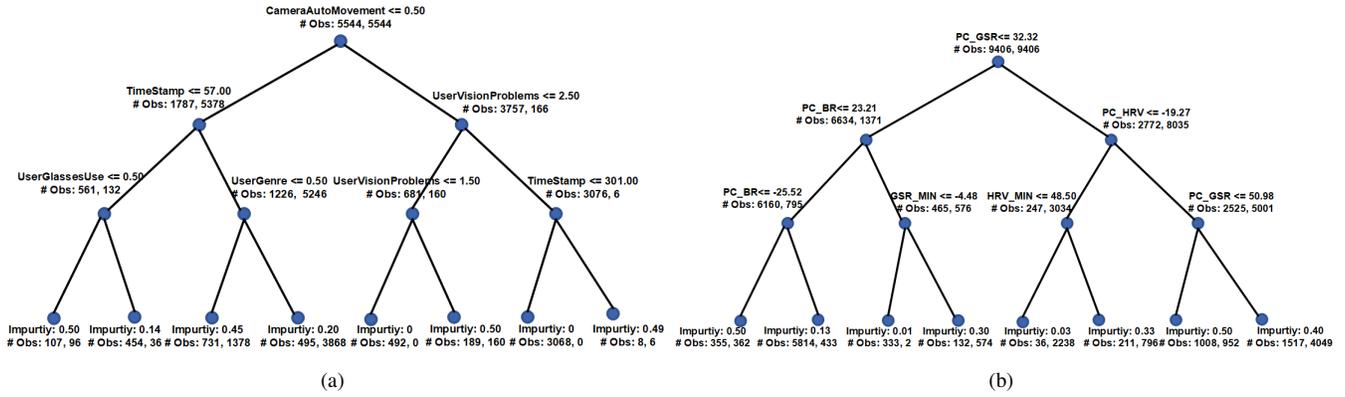

Figure 3: The global explanation of DT-based cybersickness classification. (**a**) gameplay dataset, (**b**) physiological dataset.

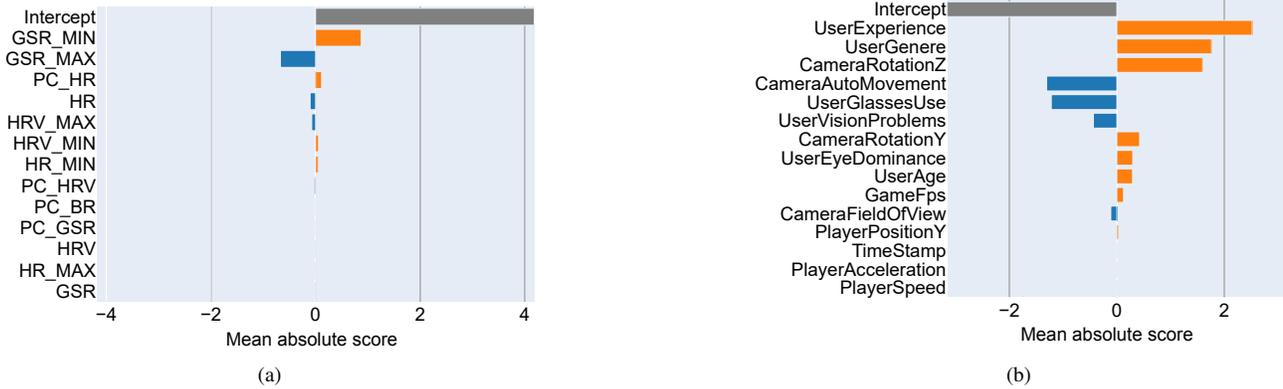

Figure 4: The global explanation of LR-based cybersickness classification. (**a**) physiological dataset, (**b**) gameplay dataset.

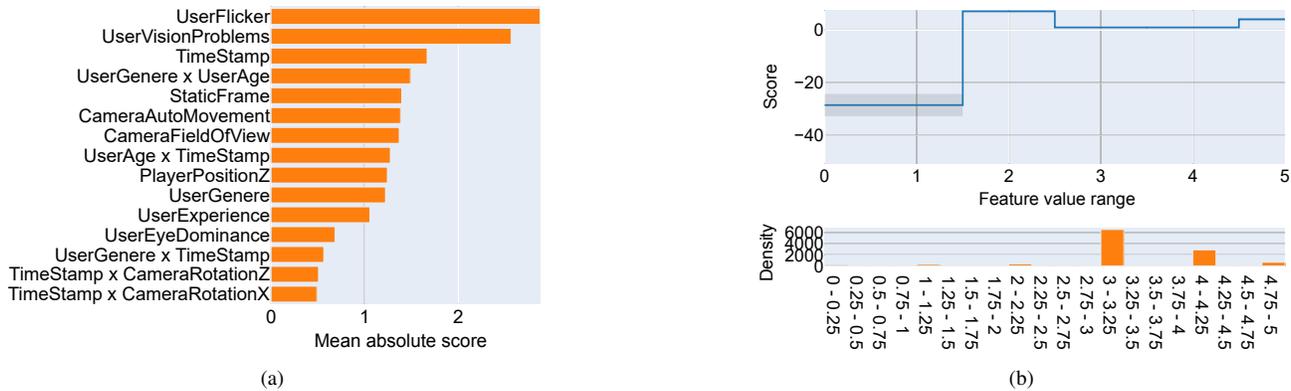

Figure 5: The global explanation of EBM-based cybersickness classification for gameplay dataset. (**a**) overall feature importance, (**b**) explanation for the feature *User vision problem*.

### 5.5 Cybersickness Regression Using Linear Regression

As shown in Table 3 and Table 4, MSE, RMSE, MAE, and ($R^2$) values for the cybersickness regression using LIR are 0.222, 0.471, 0.461 and −0.06, respectively for the physiological dataset; and 0.1414, 0.38, 0.301 and 0.081, respectively for the gameplay dataset. Note that the negative $R^2$ score indicates bad prediction, which is not obvious in the gameplay dataset.

The LIR works well on the dataset with small features and less bias. However, the physiological dataset contains a large number of continuous variables, which lead to high error in the corresponding model and hence, causes a negative $R^2$ score.

### 5.6 Cybersickness Regression using EBM

The cybersickness regression using the EBM outperforms the DT and LIR models for both the physiological and gameplay datasets. For instance, the cybersickness regression using the EBM has 44 times lower MSE than that with the DT and LIR models in the case of the physiological dataset. The gameplay dataset also has 2 times lower MSE than other ML models in this paper. Table 3 and

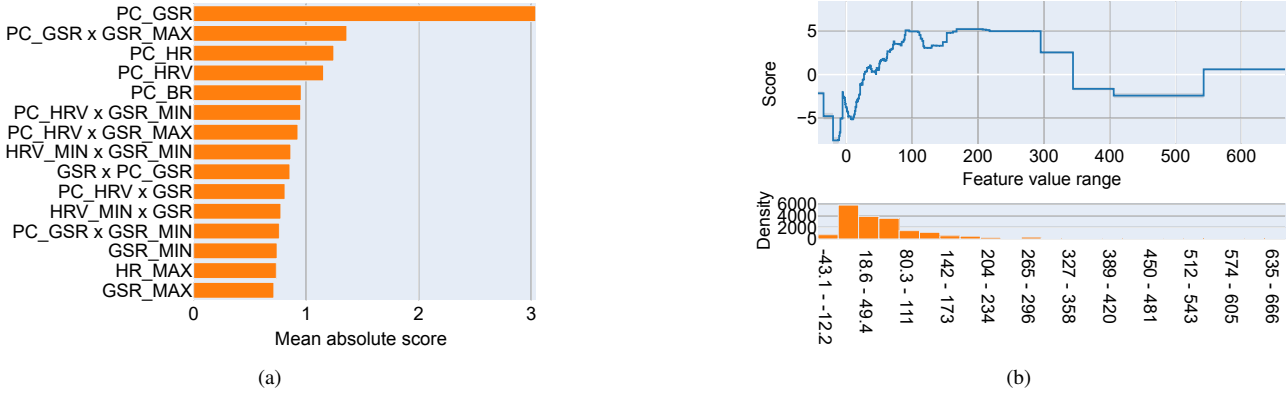

Figure 6: The global explanation of EBM-based cybersickness classification for the physiological dataset. (**a**) overall feature importance, (**b**) explanation for the feature *PC_GSR*.

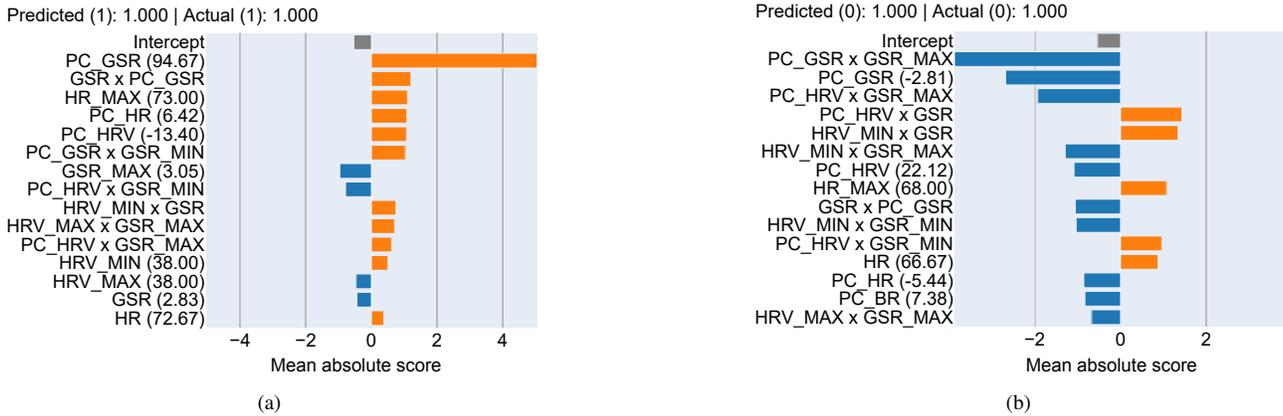

Figure 7: The local explanation of EBM-based cybersickness classification for the physiological dataset. (**a**) explanation for actual cybersickness, (**b**) explanation for no cybersickness.

Table 4 show that the cybersickness regression using EBM results in ($R^2$) score, RMSE, MAE, and MSE as 0.975, 0.071, 0.0454, and 0.005 for the physiological dataset and 0.45, 0.27, 0.071, and 0.073 for the gameplay dataset, respectively. The reason behind the poor performance of the DT and LIR models is the fact that they highly depend on the variation in the data. The higher the variations in the data, the lesser the chance of good cybersickness regression with the DT and LIR models. In contrast, as mentioned before, the EBM model is independent of the features in the data, and thus EBM exhibits a better regression outcome.

### 5.7 Cybersickness Global Explanation

This section explains the results obtained from the global explanation of the cybersickness using DT, LR, and EBM.

#### 5.7.1 Cybersickness Global Explanation Using Decision Tree

Figure 3a illustrates the global feature ranking for the gameplay dataset where the root node contains the *auto-movement of camera* as a feature for 11088 observations (#Obs). Out of these observations, (5544, 5544) sample refer to (cybersickness, no cybersickness). From this growing DT, the feature importance can be easily inferred; hence, the features that lead to better splits can be identified. For instance, the root node is split into *time stamps*, with #Obs (1787, 5378) and *user vision problems*, with #Obs (3757, 166) for two classes, under the threshold 0.50 *automovement of camera*. Next, the *time stamps* less than equal to 57 is used for the decision node and hence, split into *the usage of user glasses* with #Obs (561, 132) and *user genere* with #Obs (1226, 5246). The decision node with the feature *user glasses usage* has an impurity value of 0.50, identifies 107 'cybersickness' and 96 'no cybersickness' samples.

We can also reason about the cybersickness using this DT. For instance, if the *usage of user glasses* is high, then the user will be more likely to suffer from the cybersickness. Likewise, Figure 3b illustrates the feature importance for the physiological dataset using the growing DT. For instance, the *PC_BR* which corresponds to the percentage change of BR measurement at the decision node, indicates that the higher the value of *PC_BR*, the higher the chances of the user suffering from cybersickness. The cybersickness classification using the DT model for the physiological dataset has quite low accuracy, as discussed in Section 5.1. So, there is a wrong split of the features such as, *PC_BR* in the decision node in Figure 3b. Such misclassified cybersickness explanation provides insights into the classification results and builds the trust of the model outcome to take further decisions.

#### 5.7.2 Cybersickness Global Explanation using Logistic Regression

Figure 4 illustrates the feature importance of both the physiological and the gameplay datasets for the global explanation of the cybersickness using the LR model. In Figure 4a, the most predictive feature *GSR_MIN*, which corresponds to the minimum GSR measurement, of the cybersickness classification is highlighted with the orange color. However, the second leading feature for the no cybersickness classification *GSR_MAX* which corresponds to the maximum GSR measurement, is highlighted with the blue color. Interestingly, the feature for cybersickness and no cybersickness classification both correspond to the GSR measurement. This means that the LR model has many false positives and negatives for the physiological dataset. Hence, the cybersickness explanation using the LR model is not acceptable for further decision-making in the physiological dataset.

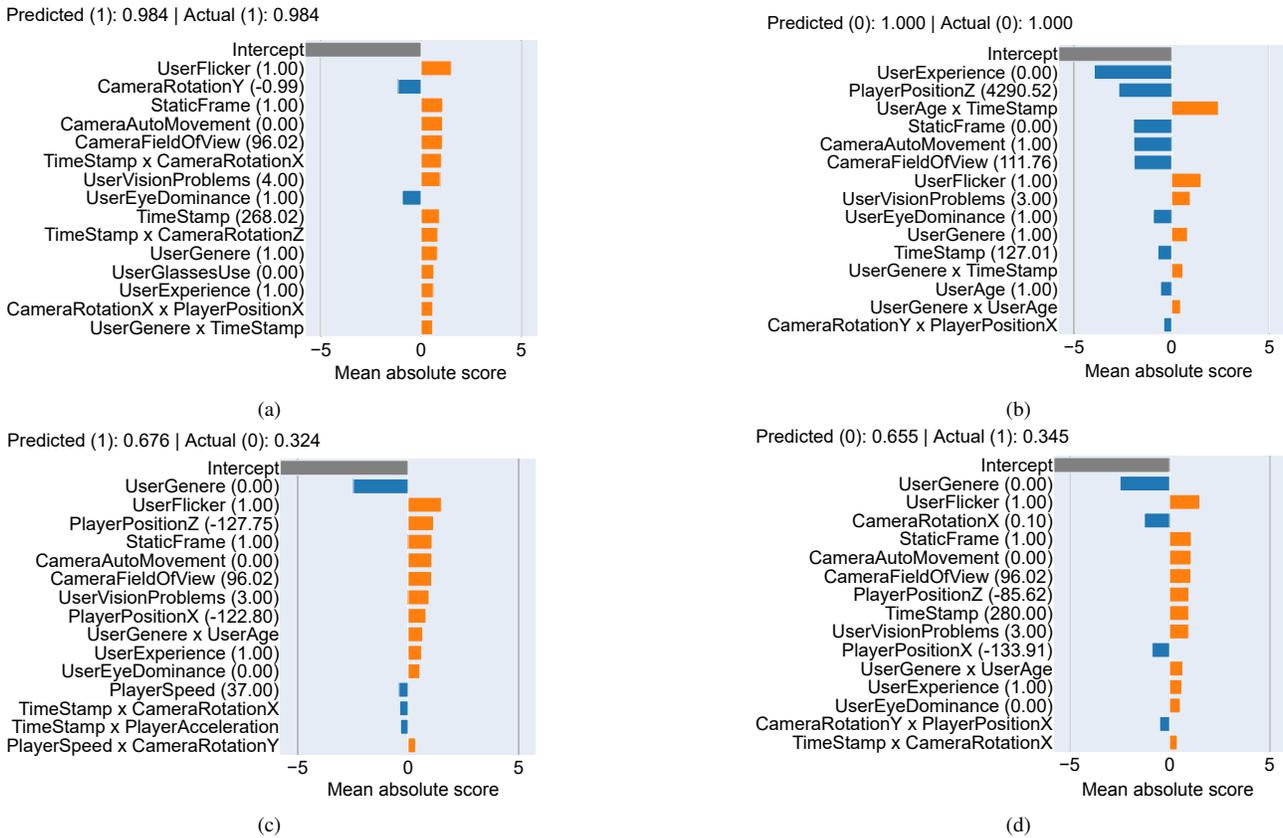

Figure 8: The local explanation of EBM based cybersickness classification for gameplay dataset (**a**) explanation for actual cybersickness, (**b**) explanation for no cybersickness, (**c**) explanation for the wrong cybersickness classification (false positive), (**d**) explanation for wrong cybersickness classification (false negative).

This analysis is aligned with the results in Section 5.2. In the gameplay dataset, the most predictive feature leading to cybersickness classification is *user experience* as highlighted in orange color in Figure 4b. Another feature *auto-movement of camera* contributes more towards no cybersickness classification in the gameplay dataset. That is why the accuracy for the cybersickness classification using the gameplay dataset is higher than that of the physiological dataset and hence, provides a better feature ranking.

### 5.7.3 Cybersickness Global Explanation using EBM

Figure 5a presents the overall feature importance in the cybersickness classification using EBM for the gameplay dataset. MAS is used to calculate the overall ranking of the most important features contributing to the cybersickness classification. It is observed that the features such as *user Flicker*, *user vision problems*, *user age*, *user gender* together, *time exposure*, etc., are the most predictive features in cybersickness classification. Figure 5b provides a deeper insights into the global explanation of the cybersickness classification for the most influential feature *user vision problem*. For instance, the feature *user vision problem* has higher importance in the cybersickness classification because there are more non-zero valued samples are sampled in its density curve. A feature sampled at zero has a very small effect on the model outcome, whereas a non-zero value has a large MAS which affects the cybersickness classification. Similarly, Figure 6a provides the overall feature importance in the cybersickness classification using EBM for the physiological dataset. We observe that the features *PC_GSR*, corresponding to the percentage of GSR measurement, and *PC_HR* corresponding to the percentage of HR measurement of the user, have a much stronger influence in the cybersickness classification. That is why the percentage of GSR and HR varies much more than other features when a user feels discomfort. It is worth mentioning that higher or lower HR indicates a person's discomfort. Simialr to Figure 5b, Figure 6b also provides a deeper insights into the global explanation of the cybersickness classification for the most influential feature *PC_GSR*. Therefore, more non-zero valued datapoints are sampled in the density curve for the feature *PC_GSR*.

### 5.8 Cybersickness Local Explanation

For a local explanation of cybersickness using EBM, the first 2 samples are taken from the physiological dataset, and 4 samples are taken from the gameplay dataset to individually explain their outcome. Figure 7a shows the results for the local explanation using EBM for the physiological dataset. The EBM does not have false positives and false negatives for cybersickness classification with the physiological dataset due to high accuracy (i.e., 99.75%) as discussed in Section 5.3. Therefore, no misclassification results are considered in this paper. Figure 7a show the cybersickness classification (true positive) outcome with a MAS of 1. The yellow and blue colored bars in Figure 7a show cybersickness and no cybersickness probabilities on that individual outcome, respectively. Most features contribute to the negative impact indicated as yellow bars; hence, an accurate decision is made for cybersickness classification. For example, in Figure 7b, the decision regarding no cybersickness classification has a MAS of 1, which indicates that most of the features contribute to the positive impact. Hence, an accurate decision is made for no cybersickness classification. Furthermore, we observe that HR and GSR are the most influential feature for cybersickness classification in the first 2 samples.

The local explanation of the classified cybersickness using the EBM model for the gameplay dataset is shown in Figure 8. In Figure 8a, the MAS for both the true and predicted classes is 0.984, respectively, for the accurate classification of cybersickness. Also, most of the features contribute to the actual cybersickness classifica-tion except *camera rotation Y* and *user eye dominance* in Figure 8a. However, Figure 8b show that the majority of the features participate in the positive outcome (no cybersickness), and their MAS is 1 for the accurate classification of no cybersickness. The *user experience*, *player position on the Z-axis*, *Camera field of view*, *Camera auto movement*, and *static frame* are the most influential features for no cybersickness classification. Both Figure 8c and Figure 8d have misclassifications for the cybersickness. For instance, Figure 8c shows no cybersickness, but the EBM model classifies cybersickness (false positive) with a MAS of 0.676 for cybersickness and 0.324 for no cybersickness. The *user flicker*, *player position on the Z-axis*, *static frame* are the most influential features for misclassifications. Consequently, Figure 8d shows the false negative of the predicted cybersickness class with a MAS of 0.345 for cybersickness and with a MAS of 0.655 for no cybersickness class. Here, the *user genere* again the most influential feature for the misclassifications. Since the DT and LR models have limited cybersickness classification results (low accuracy), their local explanation has been omitted from this paper. Their feature ranking with global explanation is discussed in Section 5.7.1 and Section 5.7.2.

## 6 DISCUSSION

Our results show that the EBM model can classify the cybersickness with an accuracy of 99.75% for the physiological and 94.10% for the gameplay dataset with explainability. The precision and recall percentage for cybersickness classification using EBM were significantly higher than the other two methods (DT and LR) for both datasets. In the regression analysis task, the EBM model also predicted the ongoing cybersickness with an RMSE value of 0.27 for the gameplay dataset and an RMSE value of 0.071 for the physiological dataset. The $R^2$ and other metrics are explained in Tables 3 and 4 for both of the datasets. To ensure the trustworthiness of the proposed cybersickness classification approach, we provide explanations in terms of global and local explanations as discussed in Section 5.7 and Section 5.8. Regarding the overall feature importance ranking, in the case of the gameplay dataset, features such as exposure time, rotation, and acceleration are the most influential features in causing cybersickness (section 5.7). Consequently, for the physiological dataset, the GSR and the HR of the user are the most influential feature in causing cybersickness. The local explanations of the individual prediction provided useful insight for each sample, which is also beneficial for the misclassification cases. However, we can easily identify the features that influence misclassification using the local explanation.

Since there is no previous work that applies explainable machine learning for detecting and explaining cybersickness, our results are not directly comparable with prior works. However, we compare our results regarding cybersickness prediction performance using the bio-physiological measurement from the prediction and detection perspective. For instance, our results show that the EBM model with bio-physiological data achieved an accuracy of 99.75% and $R^2$ value of 97.5%. In contrast, Dennison et al. [13] reported accuracy of 78% and $R^2$ values 75% for the physiological data. Consequently, Kim et al. [37], Jeong et al. [30], and Chenxin et al. [57] reported cybersickness detection accuracy of 89.16%, 94.02%, and 96.85%, respectively, using EEG/ECG signals. The most relevant to our work is by Padmanaban et al. [52] that uses a binary decision tree with video and optical features for detecting cybersickness with an accuracy of 51.94%.

Prior researchers have applied different ML, and DL models to predict cybersickness severity from bio-physiological signals and HMD's integrated sensors [25, 28]. In contrast, little research has been conducted on identifying the causes of cybersickness [25, 27, 37, 52]. However, to the best of our knowledge, to date the exists no prior work on applying explainable machine learning for cybersickness detection and explanation. Indeed, such explanations can help the developer to understand the reasons behind correct and incorrect cybersickness classification. Thus, we believe that xML-based frameworks are highly suitable for cybersickness detection, which can be further utilized for deploying effective cybersickness reduction methods.

## 7 LIMITATIONS

Although the proposed xML-based system, specifically EBM, outperformed both publicly available datasets, our approach has a few drawbacks. We performed binary classification to simplify the models for explanation purposes, e.g., to understand how different features contribute to a decision (why some samples are labeled as cybersickness or no cybersickness). However, state-of-the-art literature performed multiclass classification to classify different severity levels of cybersickness [25, 28, 37, 55]. Moreover, cybersickness can vary based on the individual parameters, VR environment, and other factors [40, 58]. The dataset that we used contained was not gender balanced. Given that gender may influence cybersickness [47, 49], we plan to conduct a further study with more diverse and gender-balanced participants.

## 8 CONCLUSION AND FUTURE WORKS

In this paper, we used xML-based models namely, EBM, DT, LR, and LIR with both local and global explanations for cybersickness classification and regression. To the best of our knowledge, this is the first work to employ xML for explaining the results of cybersickness classification and regression in VR applications. To demonstrate the effectiveness of our proposed approach, we employ two different publicly available datasets i.e., physiological and gameplay datasets. Our results show that the xML-based EBM model classifies cybersickness with an accuracy of 99.75% for the physiological dataset and 94.10% for the gameplay dataset along with explainability. Also, the EBM model predicts the ongoing cybersickness FMS score range of 0 to 10 with an RMSE of 0.071 for the physiological dataset and 0.27 for the gameplay dataset. However, cybersickness classification accuracy is comparatively smaller and prediction RMSE is relatively higher for DT, LR, and LIR models when compared to the EBM model for both datasets. In the future, we plan to extend this work for classifying cybersickness severity levels (multiclass classification) using explainable machine learning. We also plan to build a cybersickness dataset by recruiting more female participants and integrating eye-tracking, and head tracking data in a real-world VR testbed. We believe that our xML-based cybersickness classification and regression approach can also be used in the cybersickness reduction frameworks.

### ACKNOWLEDGMENTS

This material is based upon work supported by the National Science Foundation under Award Number: CNS-2114035. Any opinions, findings, and conclusions or recommendations expressed in this publication are those of the authors and do not necessarily reflect the views of the National Science Foundation.